\documentclass[epsfig,here,12pt]{article}
\pdfoutput=1
\usepackage{rotating}
\usepackage{graphicx}

\def\gray{\special{ps: 0.4 setgray}}
\def\black{\special{ps: 0.0 setgray}}

\newcommand{\draft}{
\newcount\timecount
\newcount\hours \newcount\minutes  \newcount\temp \newcount\pmhours

\hours = \time
\divide\hours by 60
\temp = \hours
\multiply\temp by 60
\minutes = \time
\advance\minutes by -\temp
\def\hour{\the\hours}
\def\minute{\ifnum\minutes<10 0\the\minutes
            \else\the\minutes\fi}
\def\clock{
\ifnum\hours=0 12:\minute\ AM
\else\ifnum\hours<12 \hour:\minute\ AM
      \else\ifnum\hours=12 12:\minute\ PM
            \else\ifnum\hours>12
                 \pmhours=\hours
                 \advance\pmhours by -12
                 \the\pmhours:\minute\ PM
                 \fi
            \fi
      \fi
\fi
}
\def\fullclock{\hour:\minute}
\begin{centering}
\gray
\begin{sideways}
\font\Hugett  =cmtt12 scaled\magstep2
{\Hugett Draft: \today,\clock}
\end{sideways}
\black
\end{centering}
\vskip -1.7cm
$\phantom{a}$
} 
\begin{document}
\newcommand {\ber} {\begin{eqnarray*}}
\newcommand {\eer} {\end{eqnarray*}}
\newcommand {\bea} {\begin{eqnarray}}
\newcommand {\eea} {\end{eqnarray}}
\newcommand {\beq} {\begin{equation}}
\newcommand {\eeq} {\end{equation}}
\newcommand {\state}[1]{\mid \! \! {#1} \rangle}
\newcommand {\eqref}[1]{(\ref {#1})}
\newcommand{\preprint}[1]{\begin{table}[t] 
           \begin{flushright}              
           \begin{large}{#1}\end{large}    
           \end{flushright}                
           \end{table}}                    
\def\Acknowledgements{\bigskip  \bigskip {\begin{center} \begin{large}
             \bf ACKNOWLEDGEMENTS \end{large}\end{center}}}

\def\Dslash{\not{\hbox{\kern-4pt $D$}}}
\def\CR{\nonumber \\ }
\def\cmp#1{{\it Comm. Math. Phys.} {\bf #1}}
\def\cqg#1{{\it Class. Quantum Grav.} {\bf #1}}
\def\pl#1{{\it Phys. Lett.} {\bf #1B}}
\def\prl#1{{\it Phys. Rev. Lett.} {\bf #1}}
\def\prd#1{{\it Phys. Rev.} {\bf D#1}}
\def\prr#1{{\it Phys. Rev.} {\bf #1}}
\def\np#1{{\it Nucl. Phys.} {\bf B#1}}
\def\ncim#1{{\it Nuovo Cimento} {\bf #1}}
\def\jmath#1{{\it J. Math. Phys.} {\bf #1}}
\def\mpl#1{{\it Mod. Phys. Lett.}{\bf A#1}}
\def\jmp#1{{\it J. Mod. Phys.}{\bf A#1}}
\def\mycomm#1{\hfill\break\strut\kern-3em{\tt ====> #1}\hfill\break}
\renewcommand{\thefootnote}{\fnsymbol{footnote}}
\begin{titlepage}
\titlepage
\begin{flushright}
WIS/06/13-May-DPPA
\\
TAUP-2967/13
\end{flushright}

\vskip 1cm
\centerline{\Large \bf Tetraquarks, their Masses and Decays in
QED$_{\bf 2}$}
\vskip 1cm
\centerline{Y. Frishman$^a$\footnote{{\tt e-mail:
yitzhak.frishman@weizmann.ac.il}}
and M. Karliner$^b$\footnote{{\tt e-mail: marek@proton.tau.ac.il}}}
\begin{center}
\em $^a$Department of Particle Physics and Astrophysics
\\Weizmann Institute of Science
\\76100 Rehovot, Israel
\end{center}
\begin{center}
\em $^b$School of Physics and Astronomy
\\Beverly and Raymond Sackler Faculty of Exact Sciences
\\Tel Aviv University, Ramat Aviv, 69978, Israel
\end{center}

\begin{abstract}

Recent observations by Belle and BESIII of charged quarkonium-like
resonances give new stimulus for theoretical investigation of exotic
hadrons in general and heavy tetraquarks in particular.
We use QED$_2$,
a confining theory, as a model for the masses and decays of tetraquarks.
Here we discuss the states $(Q\bar Q q\bar q)$ and $(Q Q \bar q \bar q)$
(and its anti-particle), where $Q$ and $q$ are two fermion flavors with
masses $M$ and $m$, so that $M > m$.  We then discuss decay modes of
these states into $(Q\bar Q)$, $(q\bar q)$, $(Q\bar q)$, $(\bar Q q)$.
It turns out that $(Q\bar Q q \bar q)$ is stable, while $(Q Q \bar q
\bar q)$ is not.

\end{abstract}
\end{titlepage}

\newpage
\section{Introduction}
From the early days of the quark model and QCD a question was raised
whether in addition to the ``ordinary" baryons built from three quarks, and
mesons containing a quark and an antiquark, hadronic spectrum
includes ``exotic" states, e.g. tetraquarks containing two quarks and two
antiquarks, see e.g.
\cite{Jaffe:1976ig,Jaffe:1976ih,Ader:1981db}. The quarks and
antiquarks inside such states might couple to unusual color
representations, for example two quarks might couple to a
{\bf 6}$_c$ and the two antiquarks to
{\bf 6}$^*_c$, etc.~\cite{Karliner:2006hf}.

 In the light quark sector it is very difficult to provide
an unambiguous experimental answer to the question whether narrow
tetraquarks exist, because the putative
light tetraquarks might mix with excited states of ordinary mesons. The
situation is somewhat easier in the heavy quark sector, thanks to
the large
mass difference between the heavy $c$ and $b$ quarks ($\equiv Q$)
and the light $u$,$d$,$s$ quarks ($\equiv q$).

Despite this, until 2003 there was no clear experimental candidate for a
tetraquark containing two heavy and two light quarks. A significant
progress occurred with the discovery of $X(3872)$ resonance \cite{X3872},
which most likely is a $\bar c c \bar q q$ tetraquark, or a $\bar D D^*$
molecule, or a mixture of the two. Further dramatic progress occurred
recently, with the discovery of manifestly-exotic charged quarkonium-like
resonances:
bottomonium-like, with quantum numbers $\bar b b \bar d u$
\cite{Karliner:2008rc},\cite{Belle:2011aa} and
charmonium-like, with quantum numbers
$\bar c c \bar d u$ \cite{Ablikim:2013zna},\cite{Liu:2013dau}.

In parallel, theoretical interest has been growing
both in $\bar Q Q \bar q q$ tetraquarks and in $Q Q \bar q \bar q$ tetraquarks.

The former have the quantum numbers of the
exotic states reported by Belle and BES, while about the latter there is no
experimental information at all. The $\bar Q Q \bar q q$ tetraquarks
are expected to be above threshold for
decay into quarkonium $\bar Q Q$ and a pion, since
ground-state
quarkonia are tightly bound and pion is very light.
On the other hand,
if the $Q$ quarks are heavy enough, $\bar Q Q \bar q q$ tetraquarks cannot decay
into two heavy-light mesons, $\bar Q q$ and
$Q \bar q$. Whether or not $c$ and/or $b$ quarks
are ``heavy enough" in this sense is still an open question.

The situation is very different for $Q Q \bar q \bar q$
tetraquarks, since there is no available channel with quarkonium and a
light meson in the final state. So if they are below the threshold
for two heavy-light
mesons, they will be stable within the strong-interactions and will only
decay weakly \cite{Manohar:1992nd},\cite{Gelman}.

In addition to being interesting in their own right,
the $Q Q \bar q \bar q$ tetraquarks are intriguing because
they are likely to have properties similar to doubly-heavy $QQq$ baryons.
This is because the two light antiquarks in a
$Q Q \bar q \bar q$ tetraquark are likely coupled to a light,
spin-zero quark-like color-triplet.

Unlike the
$Q Q \bar q \bar q$ tetraquarks, whose existence is still an open question,
$QQq$ baryons ($ccq$, $bbq$, $cbq$) {\em must} exist, as there is nothing
exotic about them. Their discovery is a very difficult
experimental challenge, but
once found, they will teach us much about the possible existence or
nonexistence of
$Q Q \bar q \bar q$ tetraquarks \cite{KNtetra}.

At present it is not known how to predict
the mass and decay width of the two types of tetraquarks from first
principles. In view of this, it is useful to examine the issue in
simplified models which hopefully incorporate some of the relevant features
of the full theory. In the present paper we study the stability of the two
types of tetraquarks in QED$_2$, which is a confining theory.

\section{Bound States in QED$_{\bf 2}$}

We start with the Lagrangian of two dimensional QED with several flavors

\beq\label{action}
{\mathcal{L}} = \sum_k \bar\psi_k (i \gamma^\mu D_\mu - m_k)\psi_k - \frac{1}{4} F_{\mu\nu}F^{\mu\nu}
\eeq
where
\bea
i D_\mu &=& i \partial_\mu - A_\mu \CR
F_{\mu\nu} &=& \partial_\mu A_\nu - \partial_\nu A_\mu \equiv \epsilon_{\mu\nu} F
\eea
$A_\mu$ is the gauge potential, $\gamma^\mu$ in two dimensions are the $\sigma$ matrices, and $k$ is the flavor index.

We now follow the treatment of Ref \cite{Ellis:1992wu}.
The treatment is simplified by transforming to bosonic variables $\chi_k$, defined via
\beq
\bar\psi_k \gamma_\mu \psi_k = -\frac{1}{\sqrt\pi} \epsilon_{\mu\nu}\partial^\nu \chi_k
\label{BosonizationEq}
\eeq
The action Eq (\ref{action}) is quadratic in the gauge potentials, and after bosonization becomes quadratic in the variable $F$.
Thus it can be integrated out, and after some rescaling of the masses (see Ref \cite{Ellis:1992wu}), we get
\beq
{\mathcal{L}} = \frac{1}{2} \sum_k (\partial_\mu \chi_k)^2 -
\frac{e^2}{2\pi} (\sum_k \chi_k)^2 + \sum_k m_k^2 \cos \sqrt{4\pi} \chi_k
\eeq
In the static case, the equations of motion are
\beq\label{motion}
\chi_k^{''} - 4\alpha (\sum_l \chi_l) -  \sqrt{4\pi} m_k^2 \sin \sqrt{4\pi} \chi_k =0
\eeq
\strut\vskip-1.0cm
\noindent
where $\alpha = \displaystyle \frac{e^2}{4\pi}$.

\vskip0.3cm
The energy density of a given state is
\bea
\epsilon &=& t + v \CR
t &=& \frac{1}{2} \sum_k \chi_k^{'2}  \CR
v &=& 2\alpha (\sum_l \chi_l)^2 + \sum_k m_k^2 (1 - \cos \sqrt{4\pi} \chi_k)
\label{EnergyDensity}
\eea
where $t$ is the kinetic part in the static case, and $v$ the potential part. For the total energy we will have to integrate over all the line.

\section{Solutions and Results}

We will look for nontrivial finite-energy solutions. These turn out to be
states composed of solitons and anti-solitons in $\chi_k$ variables.
According to the
bosonization prescription, eq.~\eqref{BosonizationEq},
the former are the fermions and the latter anti-fermions.
We take the boundary conditions at $-\infty$ to vanish, namely $\chi_k(-\infty)=0$ for all $k$. For the fermions we have
$\chi_k(\infty)=\sqrt\pi$, and for the anti-fermions $\chi_k(\infty)=-\sqrt\pi$.

From the equations of motion eq.~(\ref{motion}), we get a virial theorem: multiplying both sides by $\chi_k^{'}$, we get that
$t-v={\rm const.}$, and as it vanishes at $-\infty$, we finally get $t=v$ everywhere.

As we know \cite{Ellis:1992wu}, the only states with finite energy are those
which have zero winding number, i.e. zero net charge.

\begin{figure}[h]
\includegraphics[trim=4.4cm 2.5cm 3.1cm 2.5cm,
clip=true,
scale=0.62]{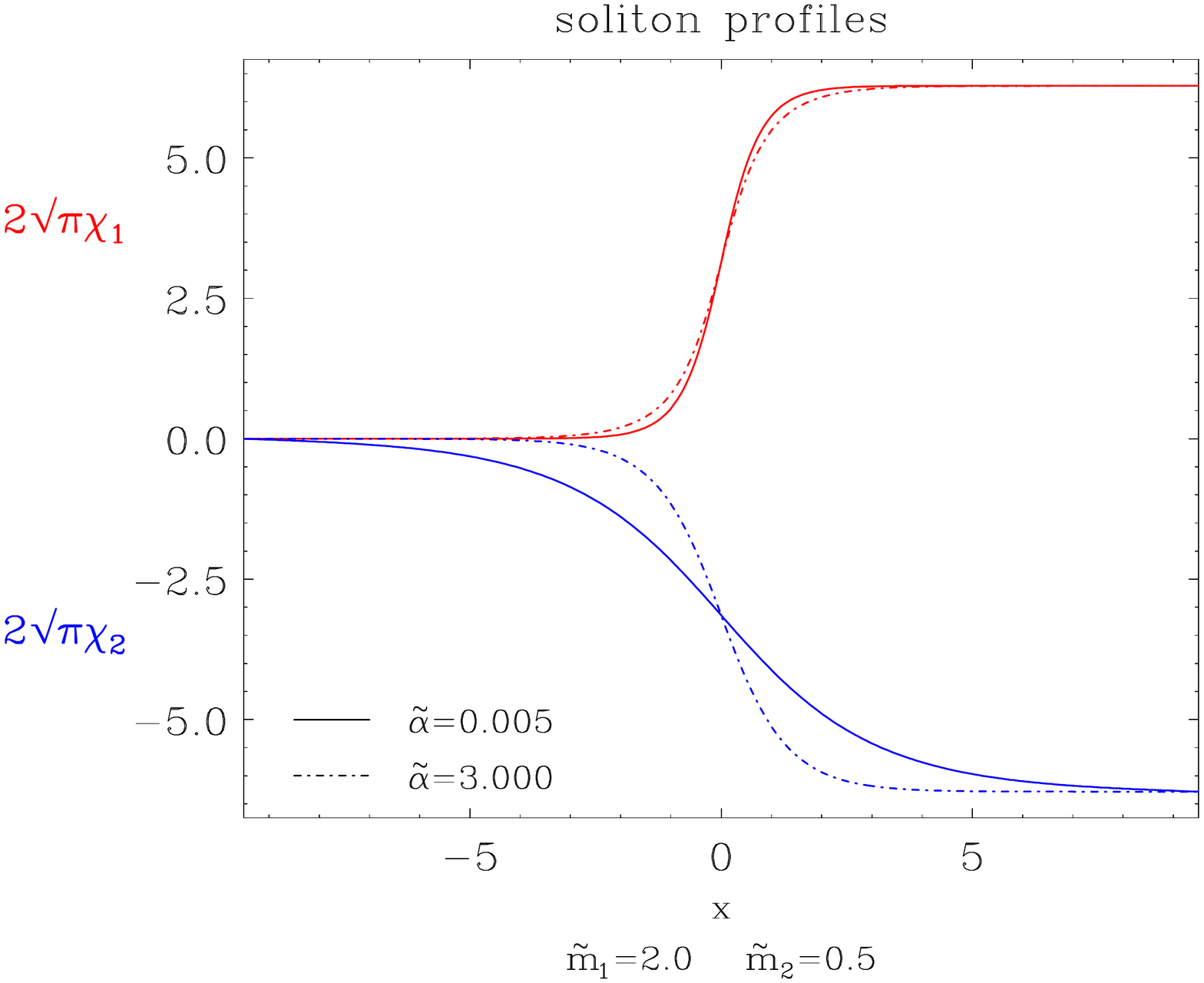}
\vskip0.3cm
\caption{\it The soliton and anti-solitons
profiles $\chi_1(x)$ and and $\chi_2(x)$
in a $(Q\bar q)$ meson,
for $\tilde m_1{=}2.0$ and $\tilde m_2{=}0.5$. Continuous lines:
$\tilde\alpha=0.005$; dash-dotted lines: $\tilde\alpha=3.$}
\label{profiles_plot}%
\end{figure}

We now focus on the case of two flavors: one heavy flavor $Q$ with mass
$M$, and one light flavor $q$ with mass m.
For free fermions
$M$ and $m$
are related to the parameters in the Lagrangian by
\bea
M &=& \frac{4}{\sqrt\pi} m_Q \CR
m &=& \frac{4}{\sqrt\pi} m_q
\eea
Equations of motion \eqref{motion} can be solved analytically for the
limiting case of equal masses, $M=m$, and numerically
for the general case $M>m$.
Sample plots of the soliton and anti-soliton profiles $\chi_1(x)$ and
$\chi_2(x)$ in a $(Q\bar q)$ meson are shown in Fig.~\ref{profiles_plot} for
$\tilde m_1=2\sqrt{\pi} m_1=2.0$, 
$\tilde m_2=2\sqrt{\pi}m_2=0.5$, 
for weak and strong coupling,
$\tilde\alpha=4\alpha = 0.005$ and 
$\tilde\alpha=4\alpha = 3.0$, respectively.

\subsubsection*{\mbox{\boldmath $(Q\bar Q q \bar q)$} tetraquarks}
Let us start with the $(Q \bar Q q \bar q)$ state.
Just like a $(Q\bar q)$ meson which is constructed from a $\chi_Q$ soliton and
a $\chi_q$ anti-soliton, a ($Q\bar Q q \bar q$) tetraquark contains
two solitons and two anti-solitons: a $Q$ soliton, a $\bar Q$ anti-soliton,
a $q$ soliton and a $\bar q$ antisoliton.

Because of the symmetry of the tetraquark under
$Q\Leftrightarrow \bar Q$,
$q\Leftrightarrow \bar q$,
for a given $\alpha$
the soliton profiles satisfy
$\chi_{\bar Q}(x) = -\chi_Q(x)$,
$\chi_{\bar q}(x) = -\chi_q(x)$. As a result, the soliton profiles cancel
each other in the term $(\sum_l \chi_l)^2$ which multiplies $\alpha$
in the expression for energy density in \eqref{EnergyDensity}.
So the mass of the state is
\def\MM{{\cal M}}
$$\MM(Q \bar Q q \bar q) = 2 M + 2m $$

Similarly,
\bea
\MM(Q \bar Q) &=& 2M   \CR
\MM(q \bar q) &=& 2m
\eea

Therefore,
\beq
\MM(Q \bar Q q \bar q) = \MM(Q \bar Q) + \MM(q \bar q)
\eeq
So there is no phase space for
$(Q\bar Q q \bar q) \to (Q \bar Q) + (q \bar q)$,
and thus no decay to this channel.

What about the decay
$(Q\bar Q q \bar q) \to (Q \bar q) + (\bar Q q)$?
We will show that this is not possible either, concluding that
in QED$_2$ the $(Q\bar Q q \bar q)$ tetraquark is stable.
To show this, we first note that the $(Q\bar Q q\bar q)$ tetraquark mass does
not depend on $\alpha$, since for any $\alpha$ we have
\beq
\MM(Q \bar Q q \bar q,\alpha)
= \MM(Q \bar q; \alpha{=}0) + \MM(\bar Q q; \alpha{=}0)
\label{QbarQqbarqMass}
\eeq

\begin{figure}[t]
\includegraphics[trim=4.3cm 2.3cm 3.5cm 2.5cm,
clip=true,
scale=0.62]{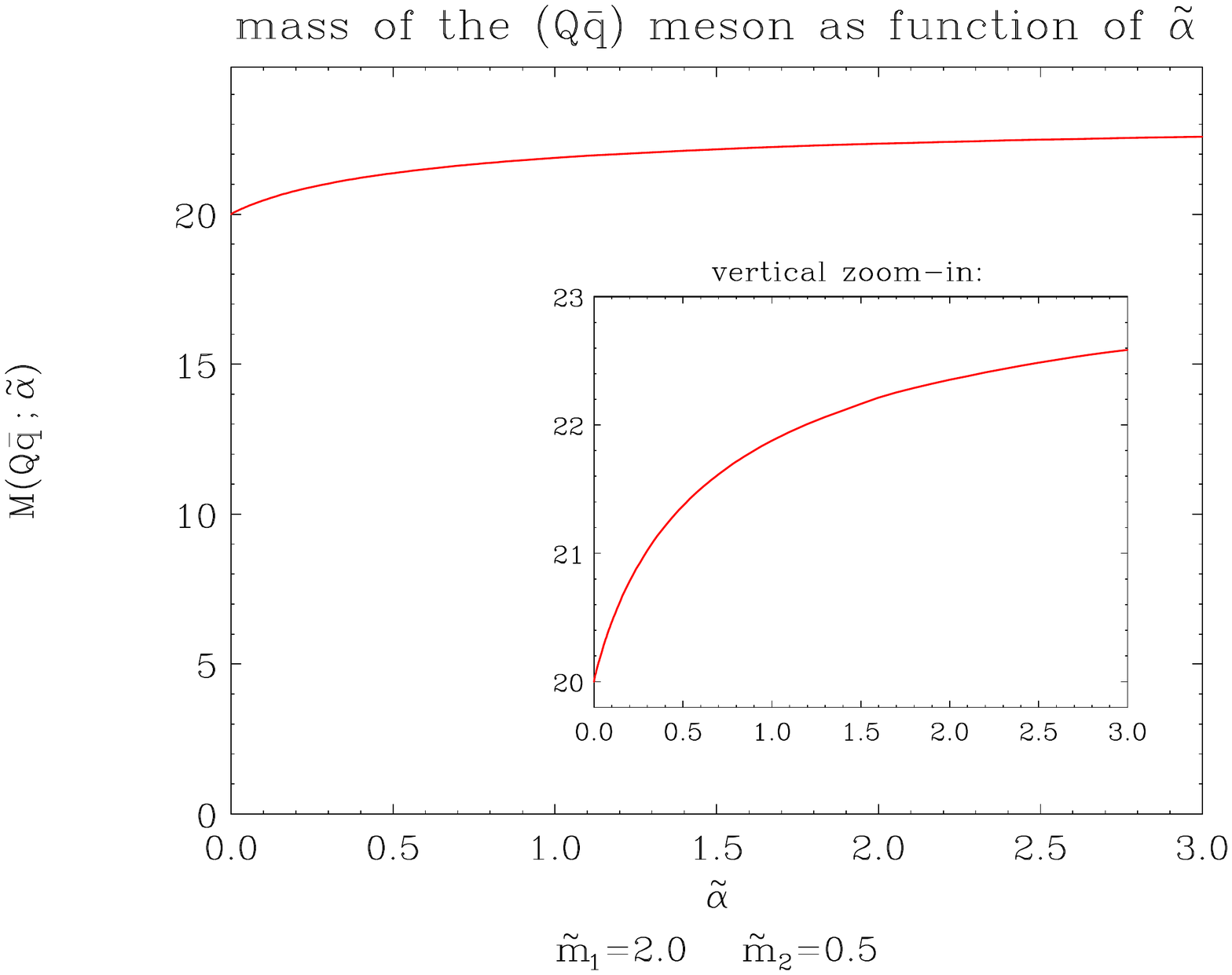}
\caption{\it Mass of the $(Q\bar q)$ meson as function of $\alpha$,
for $\tilde m_1{=}2.0$ and $\tilde m_2{=}0.5$.}
\label{Qqmass_plot}%
\end{figure}

We will now show that unlike the tetraquark mass, the $(Q\bar q)$
meson mass grows with $\alpha$, i.e. $\MM(Q\bar q)$
 is a monotonically increasing function of $\alpha$.
For this we use the Hellmann-Feynman theorem,
\beq
\frac{\partial}{\partial\alpha}M(Q, \bar q;\alpha)=
\left\langle Q\bar q\left\vert\
\frac{\partial}{\partial\alpha}\int dx\,\epsilon(x)\right\vert
Q\bar q\right\rangle
\eeq
But
\beq
\frac{\partial}{\partial\alpha}\epsilon(x) = 2 (\sum_l \chi_l(x))^2 > 0
\eeq
which proves our assertion that the $Q\bar q$
mass is monotonically increasing with $\alpha$\footnote{Except
in the limit $M \to m$, when $\chi_1(x)+\chi_2(x)=0$.}, as
demonstrated in Fig.~\ref{Qqmass_plot} for $\tilde m_1=2, \tilde m_2=0.5$.
Clearly, the same applies
to the $\bar Q q$ mass. Therefore, for any $\alpha\neq 0$ we have
\beq
\MM(Q \bar Q q \bar q,\alpha)
< \MM(Q \bar q; \alpha) + \MM(\bar Q q; \alpha)
\label{QbarQqbarqMassI}
\eeq

\subsubsection*{\mbox{\boldmath $(Q Q \bar q \bar q)$} tetraquarks}
We now consider the other type of tetraquarks,
$(Q Q \bar q \bar q)$ and in its conjugate.
From the equations of motion and the symmetry, we deduce that
\beq
\MM(Q Q \bar q \bar q; \alpha) = 2 M(Q \bar q, 2\alpha)
\eeq

The factor of 2 in front of the right hand side follows from the fact that each quark appears
twice in the tetraquark, while only once in the meson on the right, in the calculation of the
energy.
As for the factor of 2 multiplying $\alpha$ in the meson on the right, it follows from the equation of motion eq.~(\ref{motion}),
as also here each quark appears twice in the tetraquark as compared with the meson, in the term multiplying $4\alpha$.

Now, from the Hellmann-Feynman theorem,
\beq
M(Q\bar q; 2\alpha) > M(Q \bar q; \alpha)
\eeq
Thus the $(Q Q \bar q \bar q)$ tetraquark
will decay into two $(Q \bar q)$ mesons.

\section{Conclusions and Outlook}
We have studied the masses and the possible decay modes of two types of
tetra\-quarks containing two heavy and two light (anti)quarks
in QED$_2$, $(Q\bar Q q\bar q)$ and $(QQ\bar q \bar q)$.

{\em A priori}, there are two possible decay modes of $(Q\bar Q q\bar q)$:
$$(Q\bar Q q\bar q)\to (Q\bar Q) + (q\bar q)$$
and
$$(Q\bar Q q\bar q)\to (Q\bar q) + (\bar Q q).$$

We find that
regardless of the mass difference between the heavy and the light quarks,
and for all values of the coupling constant,
the $(Q\bar Q) + (q\bar q)$ final state is degenerate
 in mass with the $(Q\bar Q q\bar q)$ tetraquark,
and the $(Q\bar q) + (\bar Q q)$ state is always heavier than the
tetraquark, which is below threshold for decay into two
heavy-light mesons and therefore is stable.

For the $(QQ\bar q \bar q)$ tetraquark
the only {\em a priori} possible decay mode is
$$(QQ\bar q \bar q) \to (Q\bar q) + (Q\bar q).$$
In this case we find that
regardless of the mass difference between the heavy and the light quarks,
and for all values of the coupling constant,
the tetraquark is heavier than the two mesons and will, therefore be
unstable.

It is interesting to compare these results with QCD in four dimensions.
In QCD$_4$ the binding energy of $(Q\bar Q q \bar q)$ scales like
$\alpha_s^2 m_Q$, while the binding energy of $(Q\bar q) + (\bar Q q)$ scales
like $\Lambda_{QCD}$, so for sufficiently large $m_Q$ the
$(Q\bar Q q \bar q)$ tetraquark cannot decay to  $(Q\bar q) + (\bar Q q)$.

\section*{Acknowledgments}
The research of M.K. was supported in part by the Einstein Center for
Theoretical Physics at the Weizmann Institute.

\vfill\eject

\end{document}